\def\referee{0}
\if\referee1
	\documentclass[referee]{aa}
\else
	\documentclass{aa}
\fi
\usepackage{graphicx}
\usepackage{tabularx}
\usepackage[detect-all=true]{siunitx}
\DeclareSIUnit\astronomicalunit{au}
\DeclareSIUnit\parsec{pc}
\DeclareSIUnit\year{yr}
\usepackage[varg]{txfonts}
\usepackage{tikz}
\usetikzlibrary{calc,shapes,arrows,positioning}
\definecolor{new_green}{HTML}{2ECC40}
\usepackage{natbib, twoopt}
\usepackage[breaklinks=true, colorlinks, citecolor=black, linkcolor=black]{hyperref} 
\bibpunct{(}{)}{;}{a}{}{,}             
\makeatletter
  \newcommandtwoopt{\citeads}[3][][]{\href{http://adsabs.harvard.edu/abs/#3}%
    {\def\hyper@linkstart##1##2{}%
     \let\hyper@linkend\@empty\citealp[#1][#2]{#3}}}
  \newcommandtwoopt{\citepads}[3][][]{\href{http://adsabs.harvard.edu/abs/#3}%
    {\def\hyper@linkstart##1##2{}%
     \let\hyper@linkend\@empty\citep[#1][#2]{#3}}}
  \newcommandtwoopt{\citetads}[3][][]{\href{http://adsabs.harvard.edu/abs/#3}%
    {\def\hyper@linkstart##1##2{}%
     \let\hyper@linkend\@empty\citet[#1][#2]{#3}}}
  \newcommandtwoopt{\citeyearads}[3][][]%
    {\href{http://adsabs.harvard.edu/abs/#3}
    {\def\hyper@linkstart##1##2{}%
     \let\hyper@linkend\@empty\citeyear[#1][#2]{#3}}}
\makeatother
\begin{document}

\title{GG Tau A: Dark shadows on the ringworld}


\author{R. Brauer\inst{1,2}
	\and
	E. Pantin\inst{1}
	\and
	E. Di Folco\inst{3,4}
	\and
	E. Habart\inst{2}
	\and
	A. Dutrey\inst{3,4}
	\and
	S. Guilloteau\inst{3,4}
}

\institute{CEA Saclay - Service d'Astrophysique,
	Orme des Merisiers, Bât 709, 91191 Gif sur Yvette, France\\
	\email{[robert.brauer;eric.pantin]@cea.fr}
	\and
	Institut d'astrophysique spatiale,
	CNRS UMR 8617, Université Paris-Sud 11, Bât 121, 91405, Orsay, France\\
	\email{emilie.habart@ias.u-psud.fr}
	\and
	Univ. Bordeaux, Laboratoire d'Astrophysique de Bordeaux,
	UMR 5804, 33270 Floirac, France
	\and
	CNRS, LAB, UMR 5804, 33270 Floirac, France
}

\date{}


\abstract
{With its high complexity, large size, and close distance, the ringworld around GG Tau A is an appealing case to study the formation and evolution of protoplanetary disks around multiple star systems. However, investigations with radiative transfer models are usually neglecting the influence of the circumstellar dust around the individual stars.}
{We investigate how circumstellar disks around the stars of GG Tau A are influencing the emission that is scattered at the circumbinary disk and if constraints on these circumstellar disks can be derived.}
{We perform radiative transfer simulations with the code POLARIS to obtain spectral energy distributions and emission maps in the H-Band (near-infrared). Subsequently, we compare them with observations to achieve our aims.}
{We studied the ratio of polarized intensity at different locations in the circumbinary disk and conclude that the observed scattered-light near-infrared emission is best reproduced, if the circumbinary disk lies in the shadow of at least two co-planar circumstellar disks surrounding the central stars. This implies that the inner wall of the circumbinary disk is strongly obscured around the midplane, while the observed emission is actually dominated by the most upper disk layers. In addition, the inclined dark lane ("gap") on the western side of the circumbinary disk, which is a stable (non rotating) feature since ${\sim}\SI{20}{\year}$, can only be explained by the self-shadowing of a misaligned circumstellar disk surrounding one of the two components of the secondary close-binary star GG Tau Ab.}
{}


\maketitle
%
\section{Introduction}
\label{sect:introduction}
The formation of stars and their surrounding disks in a collapse of a molecular cloud is still an ongoing topic of research \citep[e.g.,][]{petr-gotzens_young_2015, dutrey_gg_2016}.
For instance, a significant amount of stars seem to be formed as multiple star systems with one or more additional stars \citep[e.g.,][]{duquennoy_multiplicity_1991, kraus_coevality_2009, wurster_impact_2017}. Compared to disks around a single star, these systems are able to host a common circumbinary disk as well as disks around their individual stars, which depends strongly on the separation of their stellar components \citep[e.g.,][]{dutrey_gg_2016}. Due to the various physical conditions and increased dynamical complexity of these multiple star systems, additional scientific questions have to be addressed. Main questions are how planets can be formed in either the circumbinary or the circumstellar disks and how these disks and their stars are interacting with each other via radiation and gravitation.

One of the best objects to investigate these questions is the multiple star system GG Tau A owing to its close distance and large size ($d=\SI{140}{\parsec}$, see \citealt{dutrey_gg_2016} for a recent review of the system characteristics). GG Tau A has a central binary consisting of at least two stars, and a ring-like circumbinary disk with an extent from $\SI{190}{\astronomicalunit}$ to $\SI{260}{\astronomicalunit}$ \citep[see Figs. \ref{fig:model_description:circumbinary_disk:density_distribution} and \ref{fig:model_description:circumbinary_disk:orbit_illustration} for illustration;][]{duchene_multiwavelength_2004, dutrey_gg_2016}. In addition, it is part of a wide binary with GG Tau B \citep[separation $\sim\SI{1400}{\astronomicalunit}$,][]{white_test_1999}. Many physical conditions and characteristics of the GG Tau A system were constrained by previous observations at multiple wavelengths \citep{guilloteau_gg_1999, white_test_1999,di_folco_gg_2014, dutrey_possible_2014, dutrey_gg_2016}. However, several important issues are still not fully understood. For example, with the recent finding that the star GG Tau Ab could be a close binary consisting of the stars Ab1 and Ab2, new questions with regard to the possible circumstellar disks around these stars arise \citep{di_folco_gg_2014}. However, the spatial resolution of current instruments/observatories is only sufficient to detect but not resolve them.

In this study, we perform radiative transfer (RT) simulations to derive constraints on the structure and orientation of the circumstellar disks by comparing simulated emission maps and spectral energy distributions with observations. In contrast to previous studies, we are considering the circumstellar disks as spatially resolved in our model of GG Tau A. This allows us to study the impact of the structure and orientation of the circumstellar disks on the surrounding circumbinary disk by casting shadows in the scattered light and influencing the heating of the circumbinary disk. For example, the binary system HD142527 exhibits shadows due to misaligned inner disks \citep{casassus_cooling_2019}. Even some single stars show similar effects (e.g., RXJ1604.3-2130, \citealt{pinilla_variable_2018}). We perform the radiative transfer simulations with the RT code POLARIS \citep{reissl_radiative_2016, brauer_magnetic_2017-1}, which was recently updated to consider complex structures of circumstellar disks (Brauer et al. in prep.).

We structured our study as follows. We begin with a description of the radiative transfer code that we use for our simulations (Sect. \ref{sect:polaris}). Subsequently, we introduce the model of GG Tau A, which includes the inner circumstellar disks as well as the surrounding circumbinary disk (Sect. \ref{sect:model_description}). Then, we present our results and investigate the impact of the circumstellar disks on the emission of the circumbinary disk (in Sect. \ref{sect:comparison_results}). Finally, we summarize our conclusions in Sect. \ref{conclusions}.
%
\section{The RT code POLARIS}
\label{sect:polaris}
We apply the three-dimensional continuum RT code POLARIS \citep[POLArized RadIation Simulator,][]{reissl_radiative_2016, brauer_magnetic_2017-1}. It solves the radiative transfer problem self-consistently with the Monte Carlo method. Main applications of POLARIS are the analysis of magnetic fields by simulating the polarized emission of elongated aligned dust grains and Zeeman split spectral lines. However, recent developments added circumstellar disk modeling as a new main application to POLARIS (Brauer et al. in prep.).

We use POLARIS to calculate the spatial dust temperature distribution of our GG Tau A model and calculate emission maps as well as spectral energy distributions based on the thermal emission of the dust grains and the stellar emission scattered at the dust grains. The considered dust grains are described in Sect. \ref{sect:model_description:dust}. The code was recently published on the POLARIS homepage\footnote{\url{http://www1.astrophysik.uni-kiel.de/~polaris/}}. For reference, the simulations in this study were performed with the version 4.03. of POLARIS.
%
\section{Model description}
\label{sect:model_description}
Our theoretical model of GG Tau A consists of three stars (Aa, Ab1, and Ab2), each optionally surrounded by a circumstellar (CS) disk, and a large circumbinary (CB) disk that surrounds all stars. We consider the stellar component GG Tau Ab to be a binary (Ab1 and Ab2), based on the findings of \cite{di_folco_gg_2014}. In the following, we describe each part of our radiative transfer model with their individual characteristics and parameters that we obtained from various studies.
%
\subsection{Stars}
\label{sect:model_description:stars}
We consider the stars GG Tau Aa, Ab1, and Ab2 as emitting black bodies with an effective temperature and luminosity. The effective temperature of the stars is obtained from their derived spectral type \citep{kaler_stars_1997}. For the stellar luminosities, the photospheric and accretion contributions are combined in a single luminosity \citep{hartigan_spectroscopic_2003, di_folco_gg_2014}. Furthermore, the luminosities of Ab1 and Ab2 are obtained by taking the derived luminosity for Ab from the work of \cite{hartigan_spectroscopic_2003} and distributing it to Ab1 and Ab2 via their luminosity ratio \citep[$L_\text{Ab1}/L_\text{Ab2}\sim\SI{2}{}$,][]{di_folco_gg_2014}. For the comparison with observations, we take the foreground extinction at all simulated wavelengths into account by using typical extinction curves for dust grains in the interstellar medium and the derived visual extinction $A_V=\SI{0.3}{}$ \citep{weingartner_dust_2001, duchene_multiwavelength_2004}. Even though a higher visual extinction was derived for Ab1 and Ab2, we assume that this increase is owing to local dust in the line-of-sight. The configuration of Ab1 and Ab2 in the simulations (see Fig. \ref{fig:model_description:circumbinary_disk:orbit_illustration}) corresponds to the reported orientation at the time of their discovery in winter 2012. The parameters of the stars are summarized in Table \ref{tab:stellar_parameter}.
\begin{table}
	\caption{Parameters of the stars GG Tau Aa, Ab1, and Ab2. The parameters with (*) are de-projected to obtain the correct values for the model space (see Fig. \ref{fig:model_description:circumbinary_disk:density_distribution}). The definitions of the position angle and inclination are illustrated in Fig. \ref{fig:model_description:circumbinary_disk:orbit_illustration}.}
	\label{tab:stellar_parameter}
	\centering
	\begin{tabular}{lccc}
		\multicolumn{2}{l}{Stellar parameter} & Value                     & Ref.                               \\
		\hline
		\multicolumn{4}{c}{\textit{GG Tau Aa}}                                                                 \\
		\hline
		Spectral type                         &                           & M0                           & 2   \\
		Effective temperature                 & $T_\text{Aa}$             & $\SI{3700}{K}$               & 6   \\
		Stellar luminosity                    & $L_\text{Aa}$             & $\SI{0.38}{L_\odot}$         & 3   \\
		Stellar accretion luminosity          & $L^\text{acc}_\text{Aa}$  & $\SI{0.122}{L_\odot}$        & 3   \\
		Visual extinction                     & $A_V$                     & $\SI{0.3}{}$                 & 3   \\
		\hline
		\multicolumn{4}{c}{\textit{GG Tau Ab1}}                                                                \\
		\hline
		Spectral type                         &                           & M2V                          & 2   \\
		Effective temperature                 & $T_\text{Ab1}$            & $\SI{3300}{K}$               & 6   \\
		Stellar luminosity                    & $L_\text{Ab1}$            & $\SI{0.133}{L_\odot}$        & 2,3 \\
		Stellar accretion luminosity          & $L^\text{acc}_\text{Ab1}$ & $\SI{0.053}{L_\odot}$        & 2,3 \\
		Visual extinction                     & $A_V$                     & $\SI{0.45}{}$                & 3   \\
		\hline
		\multicolumn{4}{c}{\textit{GG Tau Ab2}}                                                                \\
		\hline
		Spectral type                         &                           & M3V                          & 2   \\
		Effective temperature                 & $T_\text{Ab2}$            & $\SI{3100}{K}$               & 6   \\
		Stellar luminosity                    & $L_\text{Ab2}$            & $\SI{0.067}{L_\odot}$        & 2,3 \\
		Stellar accretion luminosity          & $L^\text{acc}_\text{Ab2}$ & $\SI{0.026}{L_\odot}$        & 2,3 \\
		Visual extinction                     & $A_V$                     & $\SI{0.45}{}$                & 4   \\
		\hline
		\multicolumn{4}{c}{\textit{Position / Distance}}                                                       \\
		\hline
		Distance to GG Tau A                  & $d$                       & $\SI{140}{pc}$               & 2,5 \\
		Inclination                           & $i$                       & $\SI{37}{\degree}$           & 4   \\
		Separation* (Aa $\rightarrow$ Ab)     & $r_\text{Aab}$            & $\SI{43}{\astronomicalunit}$ & 2   \\
		Separation* (Ab1 $\rightarrow$ Ab2)   & $r_\text{Ab12}$           & $\SI{5}{\astronomicalunit}$  & 2   \\
		Position angle*                       & $\text{PA}_\text{stars}$  & $\SI{335}{\degree}$          & 1   \\
	\end{tabular}
	\tablebib{
		(1)~\citet{yang_near-infrared_2017};
		(2) \citet{di_folco_gg_2014};
		(3) \citet{hartigan_spectroscopic_2003};
		(4) \citet{duchene_multiwavelength_2004};
		(5) \citet{white_test_1999};
		(6) \citet{kaler_stars_1997}.
	}
\end{table}
%
\subsection{Circumstellar disks}
\label{sect:model_description:circumstellar_disks}
Interferometric observations in the infrared and at millimeter wavelengths were able to detect a disk around the GG Tau Aa star with an outer radius of $R_\text{out}\sim\SI{7}{\astronomicalunit{}}$ \citep{guilloteau_gg_1999, dutrey_possible_2014}. Furthermore, in the work of \cite{di_folco_gg_2014}, they derived a maximum extent of possible CS disks around the stars Ab1 and Ab2 ($R_\text{out}\lesssim\SI{2}{\astronomicalunit{}}$). It is expected that these disks have similar density and temperature distributions as the inner part of circumstellar disks around T Tauri stars \citep{dutrey_gg_2016}. CS disks around the stars are also in agreement with the observation of \cite{yang_near-infrared_2017}, which shows clearly that dust is present around these stars (see Fig. \ref{fig:compare_obs:near_infrared_polarization_1}). Previous studies that performed RT simulations usually neglected the influence of the CS disks around the stars. The work of \cite{wood_gg_1999} partially included the influence of the CS disks in their simulations, but they were only able to use a simple approximation owing to the limited computational resources at that time. In this study, we properly include the CS disks around the stars in our radiative transfer model and investigate their influence on the emission of GG Tau A. To disentangle the influence of each individual CS disk, we use different models of GG Tau A, which cover all combinations of disks around the stars Aa, Ab1, and Ab2 (see Table \ref{tab:cs_disk_configurations}).

In absence of any measurement, we consider a density distribution for the CS disks with a radial decrease based on the work of \citet{hayashi_structure_1981} for the minimum mass solar nebular. Combined with a vertical distribution due to hydrostatic equilibrium similar to the work of \citet{shakura_black_1973}, we obtain the following equation:
\begin{equation}
	\rho_\text{disk}=\rho_0 \left( \frac{r}{R_\text{ref}} \right)^{-a} \exp\left(-\frac{1}{2}\left[\frac{z}{H(r)}\right]^2 \right).
	\label{eqn:disk}
\end{equation}
Here, $r$ is the radial distance from the central star in the disk midplane, $z$ is the distance from the midplane of the disk, $R_\text{ref}$ is a reference radius, and $H(r)$ is the scale height. The density $\rho_0$ is derived from the disk dust mass. The scale height is a function of $r$ as follows:
\begin{equation}
	H(r)=h_0 \left(\frac{r}{R_\text{ref}}\right)^{b}.
	\label{eqn:disk2}
\end{equation}
The exponents $a$ and $b$ set the radial density profile and the disk flaring, respectively (with $a=3\left[b-0.5\right]$ based on \citealt{shakura_black_1973}). The extent of the disk is constrained by the inner radius $R_\text{in}$ and the outer radius $R_\text{out}$. An overview of the considered parameters of the CS disks can be found in Table \ref{tab:cs_disk_parameter}. These parameters are chosen to be in agreement with previous studies of GG Tau A. However, parameters like the disk flaring and scale height were not previously derived, but they are strongly influencing the disk emission at $\lambda\in[\SI{5}{\micro\metre},\SI{50}{\micro\metre}]$. Therefore, we estimated these disk parameters by using individual disk models and comparing their emission at $\lambda=\SI{10}{\micro\metre}$ to observations.

\begin{table*}
	\caption{Various configurations of our GG Tau A model to investigate the influence of each individual circumstellar disk around the stars Aa, Ab1, and Ab2 on the emission of the circumbinary disk. Each star is either surrounded by its circumstellar disk (\checkmark) or has no surrounding dust (-). The CS disk with \checkmark$^i$ has an inclination of $i=\SI{90}{\degree}$ with respect to the plane of the CB disk and a position angle of the rotation axis of $\SI{270}{\degree}$.}
	\label{tab:cs_disk_configurations}
	\centering
	\begin{tabular}{clrccc}
		\# & Configuration            & CS disks around: & Aa         & Ab1        & Ab2            \\
		\hline
		1  & No circumstellar disks   &                  & -          & -          & -              \\
		2  & Disk around Aa           &                  & \checkmark & -          & -              \\
		3  & Disk around Ab1          &                  & -          & \checkmark & -              \\
		4  & Disk around Ab2          &                  & -          & -          & \checkmark     \\
		5  & Disks around Aa and Ab1  &                  & \checkmark & \checkmark & -              \\
		6  & Disks around Aa and Ab2  &                  & \checkmark & -          & \checkmark     \\
		7  & Disks around Ab1 and Ab2 &                  & -          & \checkmark & \checkmark     \\
		8  & Disks around all stars   &                  & \checkmark & \checkmark & \checkmark     \\
		\hline
		9  & Vertical disk around Ab2 &                  & \checkmark & \checkmark & \checkmark$^i$ \\
	\end{tabular}
\end{table*}

\begin{table}
	\caption{Parameters of the circumstellar disks around GG Tau Aa, Ab1, and Ab2.}
	\label{tab:cs_disk_parameter}
	\centering
	\begin{tabular}{lccc}
		\multicolumn{2}{l}{Disk parameter} & Value           & Ref.                                 \\
		\hline
		\multicolumn{4}{c}{\textit{GG Tau Aa}}                                                      \\
		\hline
		Disk dust mass                     & $M_\text{dust}$ & $\SI{1e-5}{M_\odot}$           & 1   \\
		Inner radius                       & $R_\text{in}$   & $\SI{0.07}{\astronomicalunit}$ & 2   \\
		Outer radius                       & $R_\text{out}$  & $\SI{7}{\astronomicalunit}$    & 1   \\
		Scale height                       & $h_0$           & $\SI{0.7}{\astronomicalunit}$  & 4   \\
		Characteristic radius              & $R_\text{ref}$  & $\SI{7}{\astronomicalunit}$    & 4   \\
		Radial density exponent            & $a$             & $\SI{2.1}{}$                   & 4   \\
		Disk flaring exponent              & $b$             & $\SI{1.2}{}$                   & 4   \\
		\hline
		\multicolumn{4}{c}{\textit{GG Tau Ab1 and Ab2}}                                             \\
		\hline
		Disk dust mass (each)              & $M_\text{dust}$ & $\SI{1e-7}{M_\odot}$           & 1   \\
		Inner radius                       & $R_\text{in}$   & $\SI{0.07}{\astronomicalunit}$ & 2   \\
		Outer radius                       & $R_\text{out}$  & $\SI{2}{\astronomicalunit}$    & 1,2 \\
		Scale height                       & $h_0$           & $\SI{0.3}{\astronomicalunit}$  & 4   \\
		Characteristic radius              & $R_\text{ref}$  & $\SI{2}{\astronomicalunit}$    & 4   \\
		Radial density exponent            & $a$             & $\SI{2.4}{}$                   & 4   \\
		Disk flaring exponent              & $b$             & $\SI{1.3}{}$                   & 4   \\
	\end{tabular}
	\tablebib{
		(1)~\citet{dutrey_gg_2016};
		(2) \citet{di_folco_gg_2014};
		(3) \citet{duchene_multiwavelength_2004};
		(4) Model fitting to flux values of \citet{mccabe_nicmos_2002}.
	}
\end{table}
%
\subsection{Circumbinary disk}
\label{sect:model_description:circumbinary_disk}
Observations of the CB disk around the stars of GG Tau A show an almost smooth and continuous ring-like dust distribution with brightness variations that are usually caused by the PSF subtraction, the scattering phase function of the dust grains, and the inclination of the disk (e.g., Fig. \ref{fig:compare_obs:near_infrared_polarization_1} \textit{top left}; \citealt{duchene_multiwavelength_2004, yang_near-infrared_2017}). Therefore, we approximate the CB disk with an azimuthally symmetric disk density distribution as shown in Eq. \ref{eqn:disk}. The parameters of the density distribution are mainly based on the findings of \cite{duchene_multiwavelength_2004}, who included an exponential decrease for the inner edge between $\SI{180}{\astronomicalunit}$ and $\SI{190}{\astronomicalunit}$ as follows:
\begin{equation}
	\rho_\text{disk+edge} =
	\begin{cases}
		\rho_\text{disk},                                                                          & \SI{190}{\astronomicalunit} \leq r < \SI{260}{\astronomicalunit}, \\
		\rho_\text{disk} \cdot e^{\frac{\left(r - \SI{190}{\astronomicalunit}\right)^2}{2 R^2_c}}, & \SI{180}{\astronomicalunit} \leq r < \SI{190}{\astronomicalunit}.
	\end{cases}
\end{equation}
The parameters that we considered for the CB disk in our model are summarized in Table \ref{tab:cb_disk_parameter}. The density distribution of the GG Tau A model including the CB disk and the (optional) three CS disks is shown in Fig. \ref{fig:model_description:circumbinary_disk:density_distribution}. We are not considering dust in regions other than the CS and CB disks, because significant dust outside the disks was not detected with ALMA observations \citep{dutrey_gg_2016}.

The sketch in Fig. \ref{fig:model_description:circumbinary_disk:orbit_illustration} illustrates the position angles of the stellar separation and the rotation axis of the CB disk inclination. The inner CS disks and the orbit of the stars are in the same plane as the CB disk, unless otherwise noted.
\begin{table}
	\caption{Parameters of the circumbinary disk around GG Tau A. The definition of the position angle is illustrated in Fig. \ref{fig:model_description:circumbinary_disk:orbit_illustration}.}
	\label{tab:cb_disk_parameter}
	\centering
	\begin{tabular}{lccc}
		\multicolumn{2}{l}{Disk parameter} & Value                 & Ref.                                  \\
		\hline
		Disk dust mass                     & $M_\text{dust}$       & $\SI{1.3e-3}{M_\odot}$        & 2,4   \\
		Inner radius                       & $R_\text{in}$         & $\SI{180}{\astronomicalunit}$ & 4     \\
		Outer radius                       & $R_\text{out}$        & $\SI{260}{\astronomicalunit}$ & 2,4   \\
		Scale height                       & $h_0$                 & $\SI{21}{\astronomicalunit}$  & 3,4,5 \\
		Characteristic radius              & $R_\text{ref}$        & $\SI{180}{\astronomicalunit}$ & 3,4,5 \\
		Inner edge cutoff                  & $R_c$                 & $\SI{2}{\astronomicalunit}$   & 4     \\
		Surface density exp.               & $a$                   & $\SI{-1.7}{}$                 & 4     \\
		Disk flaring                       & $b$                   & $\SI{1.05}{}$                 &       \\
		Position angle                     & $\text{PA}_\text{CB}$ & $\SI{277}{\degree}$           & 1     \\
	\end{tabular}
	\tablebib{
		(1)~\citet{yang_near-infrared_2017};
		(2) \citet{dutrey_gg_2016};
		(3) \citet{dutrey_possible_2014};
		(4) \citet{duchene_multiwavelength_2004};
		(5) \citet{mccabe_nicmos_2002}.
	}
\end{table}
\begin{figure}
	\resizebox{\hsize}{!}{\includegraphics{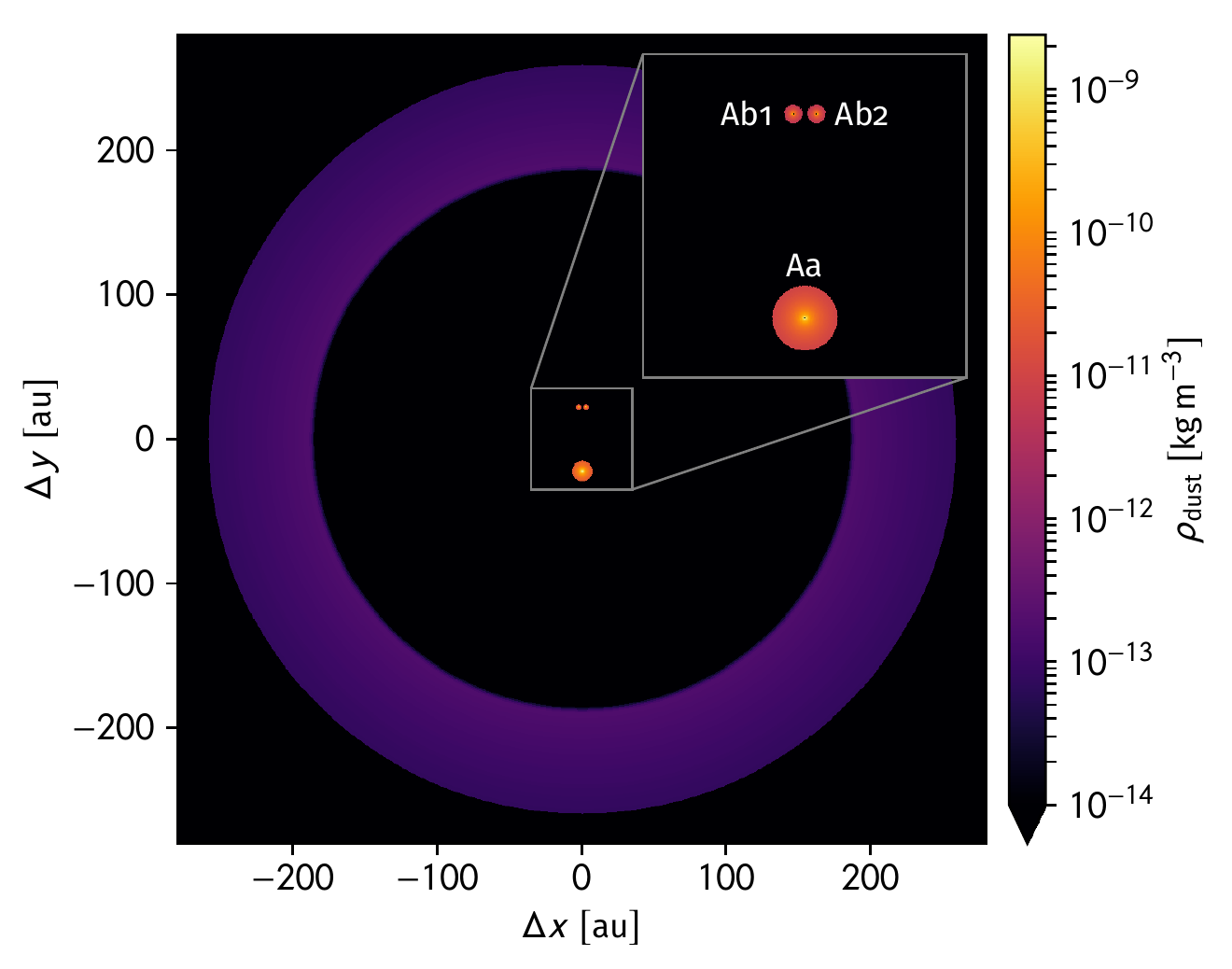}}
	\caption{Dust mass density distribution as a cut through the midplane of the GG Tau A model. The box shows the magnified inner region of the model. The stars Aa, Ab1, and Ab2 are labeled and located in the center of their respective circumstellar disks.}
	\label{fig:model_description:circumbinary_disk:density_distribution}
\end{figure}
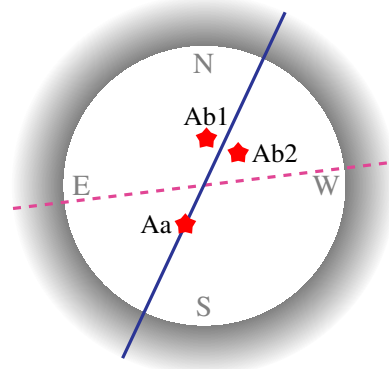
\begin{figure}
	\centering

\definecolor{new_green}{HTML}{2b3595}
\definecolor{new_purple}{HTML}{e14594}
\tikzstyle{stars} = [star, star points=5, fill=red, inner sep=0pt, minimum height=0.3cm]

\tikzset{
  ring shading/.code args={from #1 at #2 to #3 at #4}{
    \def\colin{#1}
    \def\radin{#2}
    \def\colout{#3}
    \def\radout{#4}
    \pgfmathsetmacro{\proportion}{\radin/\radout}
    \pgfmathsetmacro{\outer}{.8818cm}
    \pgfmathsetmacro{\inner}{.8818cm*\proportion}
    \pgfmathsetmacro{\innerlow}{\inner-0.01pt}
    \pgfdeclareradialshading{ring}{\pgfpoint{0cm}{0cm}}%
    {
      color(0pt)=(white);
      color(\innerlow)=(white);
      color(\inner)=(#1);
      color(\outer)=(#3)
    }
    \pgfkeysalso{/tikz/shading=ring}
  },
}

\begin{tikzpicture}[scale=2.3, thick]
	\shade[even odd rule,ring shading={from gray at 0.8cm to white at 1.1cm}]
  		(0,0) circle (0.8cm) circle (1.1cm);
  
  \node[gray, font=\large] at (0, 0.7cm) {N};
  \node[gray, font=\large] at (0, -0.7cm) {S};
  \node[gray, font=\large] at (-0.7cm, 0) {E};
  \node[gray, font=\large] at (0.7cm, 0) {W};

	\draw[very thick, rotate=277, new_purple, dashed] (0, -1.1cm) -- (0, 1.1cm);
	\node[anchor=center, new_purple] at (0, 1.2cm) {$\text{PA}_\text{CB}=\SI{277}{\degree}$ (inclination)};

	\draw[very thick, rotate=335, new_green] (0, -1.1cm) -- (0, 1.1cm);
	\node[anchor=center, new_green] at (0, 1.4cm) {$\text{PA}_\text{stars}=\SI{335}{\degree}$ (major axis of the stars)};

	\node[stars] at (0.25 * 0.423cm + 0.1 * 0.906cm, 0.25 * 0.906cm - 0.1 * 0.423cm) {};
	\node[anchor=west, xshift=0.05cm] at (0.25 * 0.423cm + 0.1 * 0.906cm, 0.25 * 0.906cm - 0.1 * 0.423cm) {Ab2};
	\node[stars] at (0.25 * 0.423cm - 0.1 * 0.906cm, 0.25 * 0.906cm + 0.1 * 0.423cm) {};
	\node[anchor=south, yshift=0.05cm] at (0.25 * 0.423cm - 0.1 * 0.906cm, 0.25 * 0.906cm + 0.1 * 0.423cm) {Ab1};
	\node[stars] at (-0.25 * 0.423cm, -0.25 * 0.906cm) {};
	\node[anchor=east, xshift=-0.05cm] at (-0.25 * 0.423cm, -0.25 * 0.906cm) {Aa};
\end{tikzpicture}
	\caption{
		Sketch of the orientations. The separations are not to scale. The blue solid line refers to the position angle of the axis from GG Tau Aa to Ab1 and Ab2 and the pink dashed line refers to the major axis of the inclined CB disk. The position angles are celestial positive ($\text{N}\rightarrow\text{E}$). Due to the inclination, the northern region of the disk is oriented towards the observer. The orientation of the Ab1/Ab2 binary corresponds to its discovery in 2012.
	}
	\label{fig:model_description:circumbinary_disk:orbit_illustration}
\end{figure}
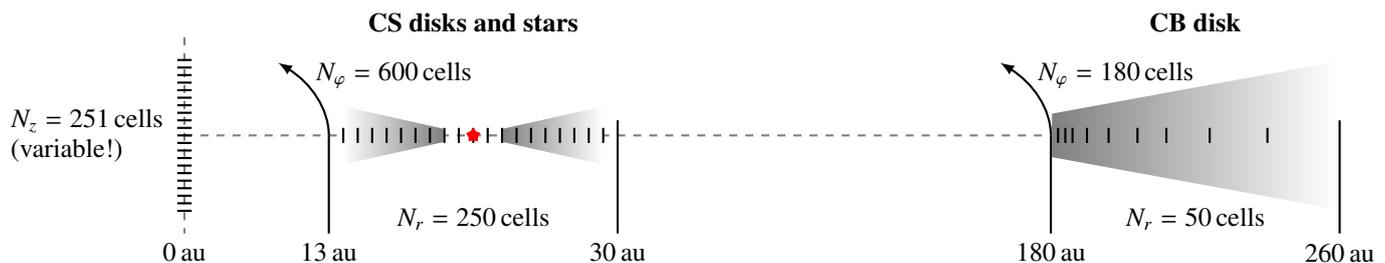
\begin{figure*}
	\centering

\definecolor{new_green}{HTML}{2b3595}
\definecolor{new_purple}{HTML}{e14594}
\tikzstyle{stars} = [star, star points=5, fill=red, inner sep=0pt, minimum height=0.2cm]

\begin{tikzpicture}[xscale=1.9, thick]
    \draw[gray, thick, dashed] (0cm, 0cm) -- (8cm, 0cm);
    \draw[gray, thick, dashed] (0cm, 1.3cm) -- (0cm, -1.3cm) node[anchor=north, black] {$\SI{0}{\astronomicalunit{}}$};
    \node[text width=2.1cm,align=left] at (-0.65cm, 0cm) {$N_z=\SI{251}{cells}$ \\ (variable!)};

    \fill[right color=gray, left color=white, draw=white] (1.8cm, 0.1cm) -- (1.1cm, 0.4cm) -- (1.1cm, -0.4cm) -- (1.8cm, -0.1cm) -- cycle;
    \fill[left color=gray, right color=white, draw=white] (2.2cm, 0.1cm) -- (2.9cm, 0.4cm) -- (2.9cm, -0.4cm) -- (2.2cm, -0.1cm) -- cycle;
    \fill[left color=gray, right color=white, draw=white] (6cm, 0.3cm) -- (8cm, 1cm) -- (8cm, -1cm) -- (6cm, -0.3cm) -- cycle;

    \node at (2cm, 1.5cm) {\textbf{CS disks and stars}};
    \node at (2cm, -1.1cm) {$N_r=\SI{250}{cells}$};

    \node at (7cm, 1.5cm) {\textbf{CB disk}};
    \node at (7cm, -1.1cm) {$N_r=\SI{50}{cells}$};

    \draw (-0.05cm, 1.0cm) -- (0.05cm, 1.0cm);
    \draw (-0.05cm, 0.9cm) -- (0.05cm, 0.9cm);
    \draw (-0.05cm, 0.8cm) -- (0.05cm, 0.8cm);
    \draw (-0.05cm, 0.7cm) -- (0.05cm, 0.7cm);
    \draw (-0.05cm, 0.6cm) -- (0.05cm, 0.6cm);
    \draw (-0.05cm, 0.5cm) -- (0.05cm, 0.5cm);
    \draw (-0.05cm, 0.4cm) -- (0.05cm, 0.4cm);
    \draw (-0.05cm, 0.3cm) -- (0.05cm, 0.3cm);
    \draw (-0.05cm, 0.2cm) -- (0.05cm, 0.2cm);
    \draw (-0.05cm, 0.1cm) -- (0.05cm, 0.1cm);
    \draw (-0.05cm, 0cm) -- (0.05cm, 0cm);
    \draw (-0.05cm, -0.1cm) -- (0.05cm, -0.1cm);
    \draw (-0.05cm, -0.2cm) -- (0.05cm, -0.2cm);
    \draw (-0.05cm, -0.3cm) -- (0.05cm, -0.3cm);
    \draw (-0.05cm, -0.4cm) -- (0.05cm, -0.4cm);
    \draw (-0.05cm, -0.5cm) -- (0.05cm, -0.5cm);
    \draw (-0.05cm, -0.6cm) -- (0.05cm, -0.6cm);
    \draw (-0.05cm, -0.7cm) -- (0.05cm, -0.7cm);
    \draw (-0.05cm, -0.8cm) -- (0.05cm, -0.8cm);
    \draw (-0.05cm, -0.9cm) -- (0.05cm, -0.9cm);
    \draw (-0.05cm, -1.0cm) -- (0.05cm, -1.0cm);

    \draw (1.1cm, 0.1cm) -- (1.1cm, -0.1cm);
    \draw (1.2cm, 0.1cm) -- (1.2cm, -0.1cm);
    \draw (1.3cm, 0.1cm) -- (1.3cm, -0.1cm);
    \draw (1.4cm, 0.1cm) -- (1.4cm, -0.1cm);
    \draw (1.5cm, 0.1cm) -- (1.5cm, -0.1cm);
    \draw (1.6cm, 0.1cm) -- (1.6cm, -0.1cm);
    \draw (1.7cm, 0.1cm) -- (1.7cm, -0.1cm);
    \draw (1.8cm, 0.1cm) -- (1.8cm, -0.1cm);
    \draw (1.9cm, 0.1cm) -- (1.9cm, -0.1cm);
    \draw (2.0cm, 0.1cm) -- (2.0cm, -0.1cm);
    \draw (2.1cm, 0.1cm) -- (2.1cm, -0.1cm);
    \draw (2.2cm, 0.1cm) -- (2.2cm, -0.1cm);
    \draw (2.3cm, 0.1cm) -- (2.3cm, -0.1cm);
    \draw (2.4cm, 0.1cm) -- (2.4cm, -0.1cm);
    \draw (2.5cm, 0.1cm) -- (2.5cm, -0.1cm);
    \draw (2.6cm, 0.1cm) -- (2.6cm, -0.1cm);
    \draw (2.7cm, 0.1cm) -- (2.7cm, -0.1cm);
    \draw (2.8cm, 0.1cm) -- (2.8cm, -0.1cm);
    \draw (2.9cm, 0.1cm) -- (2.9cm, -0.1cm);

    \draw (6.05cm, 0.1cm) -- (6.05cm, -0.1cm);
    \draw (6.1cm, 0.1cm) -- (6.1cm, -0.1cm);
    \draw (6.15cm, 0.1cm) -- (6.15cm, -0.1cm);
    \draw (6.25cm, 0.1cm) -- (6.25cm, -0.1cm);
    \draw (6.4cm, 0.1cm) -- (6.4cm, -0.1cm);
    \draw (6.6cm, 0.1cm) -- (6.6cm, -0.1cm);
    \draw (6.8cm, 0.1cm) -- (6.8cm, -0.1cm);
    \draw (7.1cm, 0.1cm) -- (7.1cm, -0.1cm);
    \draw (7.5cm, 0.1cm) -- (7.5cm, -0.1cm);

    \node[stars] at (2cm, 0cm) {};

    \draw[thick, -latex] (1cm, 0cm) arc (0:40:1.5cm);
    \node at (1.45cm, 0.8cm) {$N_\varphi=\SI{600}{cells}$};

    \draw[thick, -latex] (6cm, 0cm) arc (0:40:1.5cm);
    \node at (6.45cm, 0.8cm) {$N_\varphi=\SI{180}{cells}$};

    \draw[thick] (1cm, 0cm) -- (1cm, -1.3cm) node[anchor=north] {$\SI{13}{\astronomicalunit{}}$};
    \draw[thick] (3cm, 0.2cm) -- (3cm, -1.3cm) node[anchor=north] {$\SI{30}{\astronomicalunit{}}$};
    \draw[thick] (6cm, 0cm) -- (6cm, -1.3cm) node[anchor=north] {$\SI{180}{\astronomicalunit{}}$};
    \draw[thick] (8cm, 0.2cm) -- (8cm, -1.3cm) node[anchor=north] {$\SI{260}{\astronomicalunit{}}$};
\end{tikzpicture}
	\caption{
		Distribution of grid cells to resolve the circumstellar and circumbinary disks with our numerical model of GG Tau A. The grid is created with cylindrical geometry (radial: $r$, azimuthal: $\varphi$, vertical: $z$). The cells in $z$-direction have a variable vertical size to fit to the local density distribution (see Sect. \ref{sect:model_description:grid_resolution}). Even if only the circumstellar disk around GG Tau Aa is shown, the same distribution of cells applies to the disks around GG Tau Ab1 and Ab2.
	}
	\label{fig:model_description:circumbinary_disk:grid_illustration}
\end{figure*}
%
\subsection{Dust grains}
\label{sect:model_description:dust}
The material around the GG Tau A stars is composed of gas and dust whereby the dust grains are mainly responsible for the emission and extinction of radiation. For our simulations, we assume compact, homogeneous and spherical dust grains that consist of 62.5\% silicate and 37.5\% graphite (MRN-dust, \citealt{mathis_size_1977}; optical properties from \citealt{weingartner_dust_2001}). Furthermore, we consider the following size distribution of the dust grains \citep{mathis_size_1977}:
\begin{equation}
	\text{d}n(a)\propto a^p \text{d}a, \quad a_\text{min} < a < a_\text{max}.
	\label{eqn:dust}
\end{equation}
Here, $n(a)$ is the number of dust particles with a specific dust grain radius $a$. For the CB disk, we assume grain size limits that take the slightly larger grains ($a_\text{max}=\SI{0.5}{\micro\metre}$) into account, which are proposed by the best fit model from the work of \cite{duchene_multiwavelength_2004}. However, they obtained these limits from observations in the H-band, which are therefore only valid for dust grains in a certain vertical distance to the midplane. Nevertheless, to avoid additional free parameters, we consider these grain size limits for the whole CB disk.

At optical to near-IR wavelengths, the important (not optically thick) line-of-sights from the star to the CB disk will pass through the upper layers of the CS disks. Even if grain growth is present in the midplane of the CS disks, the upper layers should still be dominated by small grains. Therefore, we assume a single grain size limit similar to the ISM for the CS disks. An overview of the considered dust grain parameters is shown in Table \ref{tab:dust_parameter}.
\begin{table}
	\caption{Parameters of the considered dust grains.}
	\label{tab:dust_parameter}
	\centering
	\begin{tabular}{lccc}
		\multicolumn{2}{l}{Dust parameter} & Value                    & Ref.                        \\
		\hline
		Composition                        &                          & MRN-dust                & 1 \\
		Minimum size (all disks)           & $a_\text{min}$           & $\SI{5}{\nano\metre}$   & 1 \\
		Maximum size (CS disks)            & $a^\text{CS}_\text{max}$ & $\SI{250}{\nano\metre}$ & 2 \\
		Maximum size (CB disk)             & $a^\text{CB}_\text{max}$ & $\SI{500}{\nano\metre}$ & 1 \\
		Size exponent                      & $p$                      & $\SI{-3.5}{}$           & 1 \\
	\end{tabular}
	\tablebib{
		(1)~\citet{duchene_multiwavelength_2004};
		(2) \citet{mathis_size_1977}.
	}
\end{table}
%
\subsection{Grid resolution}\label{sect:model_description:grid_resolution}
In our radiative transfer simulations, we fully consider the CS disks in addition to the large outer CB disk. This requires a numerical model, i.e., grid that resolves the disk structures on different scales (from $\sim\SI{0.01}{\astronomicalunit}$ to $\sim\SI{10}{\astronomicalunit}$). We realize this by using a cylindrical grid and adjusting the grid resolution according to the position in the model. With this approach, we are able to resolve the small scale structures of the circumstellar disks and are still not excessively resolving the large scale structure of the CB disk. In the following, we describe the resolution of our grid along its three cylindrical axis, which is also illustrated in Fig. \ref{fig:model_description:circumbinary_disk:grid_illustration}.

The radial cell borders are linearly spaced with $250$ cells between $\SI{13}{\astronomicalunit}$ and $\SI{30}{\astronomicalunit}$ for the CS disks ($\Delta r\sim\SI{0.07}{\astronomicalunit}$) and $50$ cells that are exponentially distributed between $\SI{180}{\astronomicalunit}$ and $\SI{260}{\astronomicalunit}$ for the CB disk ($\Delta r\in[\SI{0.7}{\astronomicalunit},\SI{3}{\astronomicalunit}]$). In the azimuthal direction, we use two different resolutions. For the CS disks, we use $600$ cells in azimuthal direction ($\Delta a\in[\SI{0.13}{\astronomicalunit},\SI{0.3}{\astronomicalunit}]$) and for the CB disk, we use $180$ cells in azimuthal direction ($\Delta a\in[\SI{6}{\astronomicalunit},\SI{9}{\astronomicalunit}]$). In the vertical direction, we distribute $251$ cells with a vertical extent of $\SI{3}{\astronomicalunit}$ for the CS disks ($\Delta z\sim\SI{0.012}{\astronomicalunit}$). For the CB disk, we distribute the $251$ cells in vertical direction to include $10$ times the scale height (see Eq. \ref{eqn:disk2}) of the disk at each radial distance ($\Delta z\in[\SI{0.8}{\astronomicalunit},\SI{1.2}{\astronomicalunit}]$).

The center of our cylindrical grid is chosen to be in the middle between the stars Aa and Ab (middle between Ab1 and Ab2). This center point is well suited to provide a similar grid resolution to all CS disks and is still close to the barycenter of the stars. Nevertheless, other publications may consider different reference points (e.g., centered on Aa), which has to be taken into account when comparing our findings to their work.
%
\section{Comparison with observations}
\label{sect:comparison_results}
In the following sections, we compare the simulated emission maps and spectral energy distributions of our model with observations. Furthermore, we investigate the influence of the CS disks around the stars GG Tau Aa, Ab1, and Ab2 on the emission of the surrounding CB disk.
%
\subsection{Spectral energy distributions}
\label{sect:comparison_results:sed}
We simulate the spectral energy distribution of our GG Tau A models and compare them with photometric observations that were performed from visual to millimeter wavelengths. In particular, these models consist of the CB disk, the stars Aa, Ab1, and Ab2, and different combinations of the CS disks around these stars (see Table \ref{tab:cs_disk_configurations}).

\begin{figure*}
	\centering
	\includegraphics[width=\hsize]{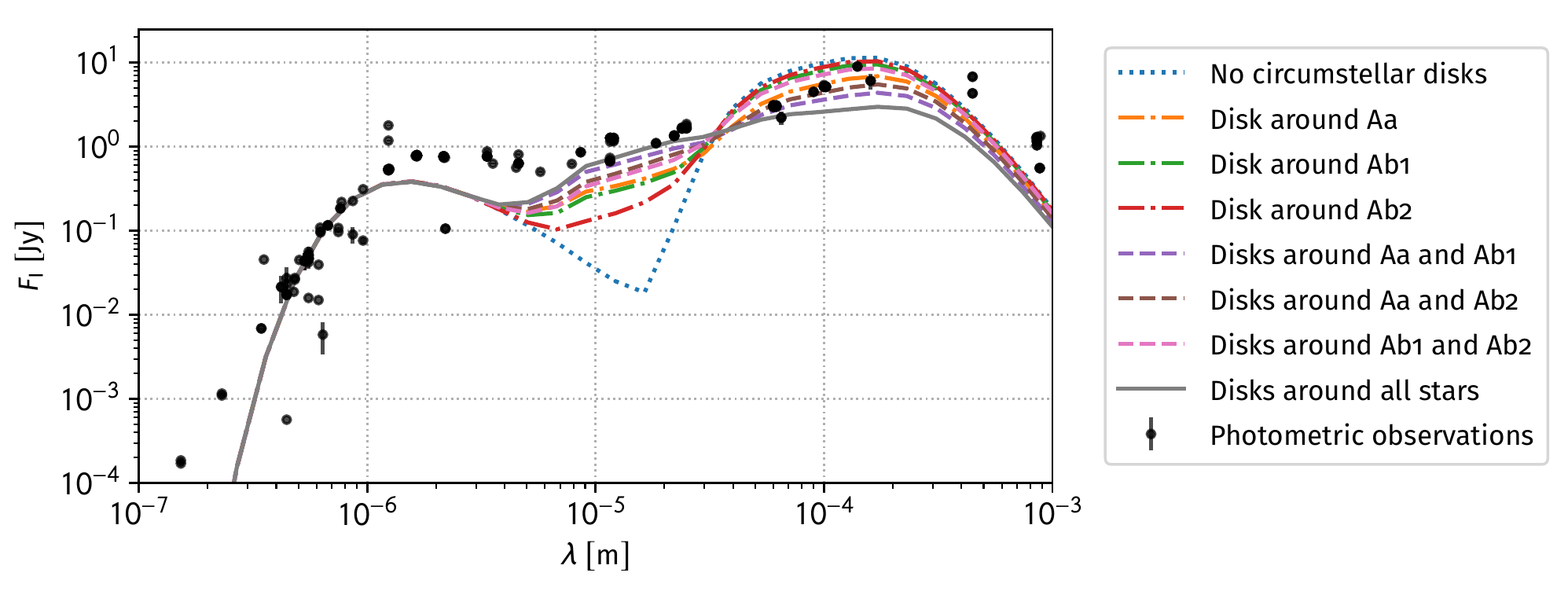}
	\caption{
		Spectral energy distribution of the GG Tau A disk model. Each energy distribution considers a different combination of the circumstellar disks, but always takes the emission of the circumbinary disk as well as the emission from all three stars into account (see Table \ref{tab:cs_disk_configurations}). For the comparison with observations, the foreground reddening was applied to the simulated spectral energy distributions \citep[$A_\text{V}=0.3$,][]{hartigan_spectroscopic_2003}. The black dots are photometric measurements obtained from the VizieR catalog service (see Appendix \ref{app:sed}).
	}
	\label{fig:comparison_results:sed:all_disks}
\end{figure*}

As illustrated in Fig. \ref{fig:comparison_results:sed:all_disks}, the CS disks are dominating the emission around $\lambda=\SI{10}{\micro\metre}$ and are necessary to obtain a level of emission as measured in observations. The underestimation of the observed photometric flux between $\lambda=\SI{1}{\micro\metre}$ and $\lambda=\SI{10}{\micro\metre}$ can be related to the usually higher dust temperatures of CS disks in multiple star systems due to accretion shocks from streamers and the stochastically heated emission of probably present nanometer-sized dust grains, as neither of them is taken into account in our simulations \citep{draine_infrared_2001,dutrey_gg_2016}. It is also related to the well known missing near-infrared flux problem, that motivated the proposition of puffed-up inner rim models, but this more complicated structure is not taken into account due to the limitations of our grid. Furthermore, the CS disks have also a significant impact on the heating of the CB disk and, therefore, on its (sub-)mm emission. From no CS disk to all CS disks, this emission decreases by a factor of up to five (e.g., $T_\text{dust}=\SI{15}{K}\rightarrow \SI{10}{K}$ at $\SI{250}{\astronomicalunit}$). Nevertheless, the slope in the $\si{\milli\metre}$ range of our simulated spectral energy distributions is steeper than in the observations, which is caused by our assumption of a single grain size distribution of relatively small dust grains in the whole CB disk. This assumption is sufficient for our simulations in the near-IR, but these dust grains underestimate the emission in the $\si{\milli\metre}$ range.

In summary, the impact of the CS disks around the stars in multiple star systems needs to be taken into account when deriving constraints from the infrared to (sub)mm wavelengths range of observed spectral energy distributions.

%

\subsection{Near-infrared emission maps}
\label{sect:comparison_results:near_ir_obs}
\begin{figure*}
	\centering
	\input{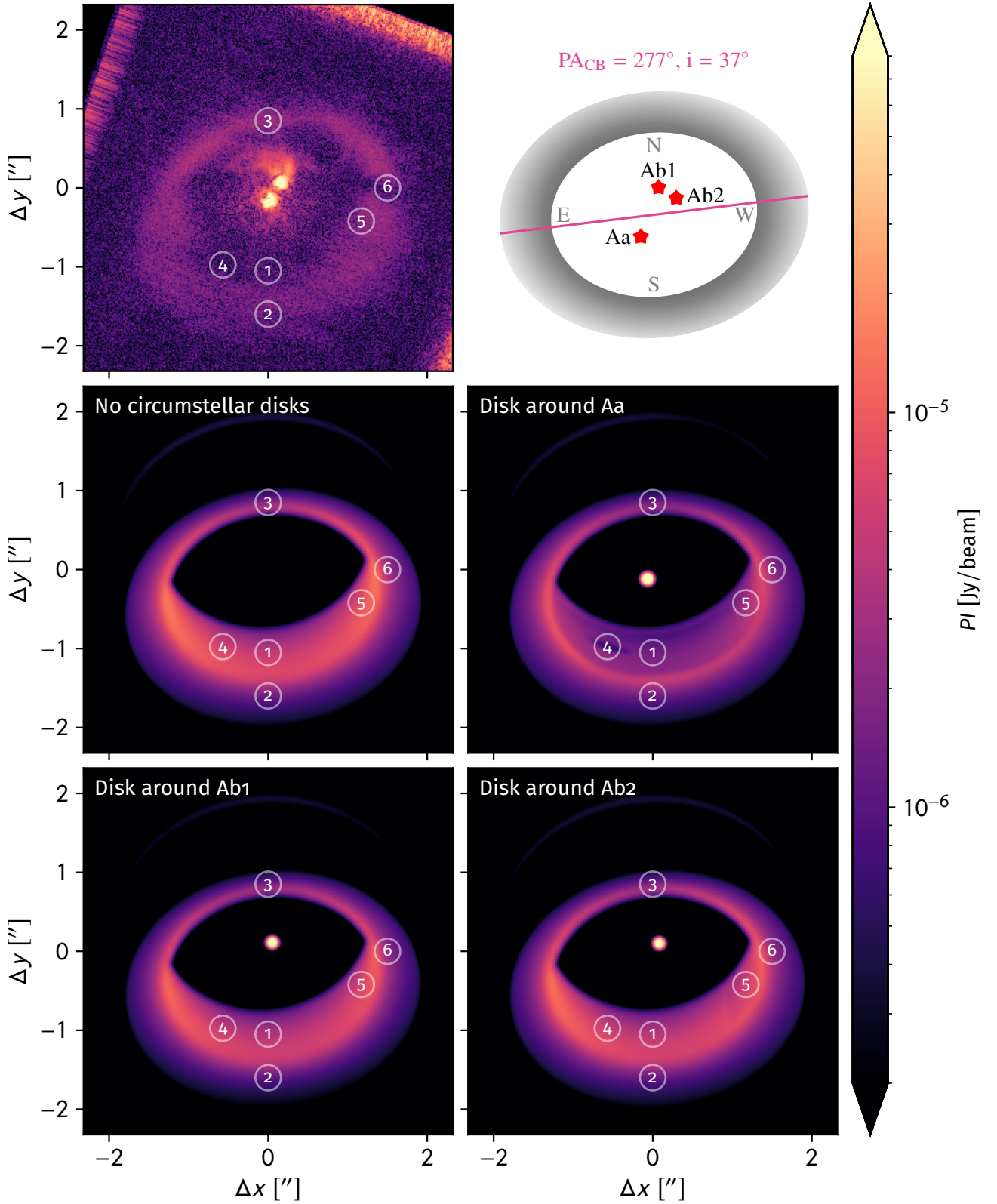}
	\caption{\textit{Top left:} Polarized intensity observation of GG Tau A in the near-infrared (H-Band) with Subaru/HiCIAO \citep{yang_near-infrared_2017}. The uncalibrated observation was fitted visually to the simulated emission maps by assuming a linear photon count. \textit{Other images:} Polarized intensity maps of GG Tau A that are simulated at $\lambda=\SI{1.65}{\micro\metre}$. The images are convolved with a gaussian beam of $\SI{0.07}{\arcsecond}\times\SI{0.07}{\arcsecond}$ to take the resolution of the Subaru/HiCIAO telescope into account. Each simulated map considers a different combination of circumstellar disks around GG Tau Aa, Ab1, and Ab2 to investigate their individual influence on the emission of the circumbinary disk (see Table \ref{tab:cs_disk_configurations}). The circles mark regions that are used to derive characteristic flux ratios from the average emission in each circle. Regions 2, 3 and 6 are the upper layers of the CB disk, whereas regions 1, 4 and 5 are located on the midplane, in the vertical wall of the circumbinary disk. Regions 4, 5 and 6 represent specific locations of projected shadows by individual CS disks.}
	\label{fig:compare_obs:near_infrared_polarization_1}
\end{figure*}

\begin{figure*}
	\centering
	\input{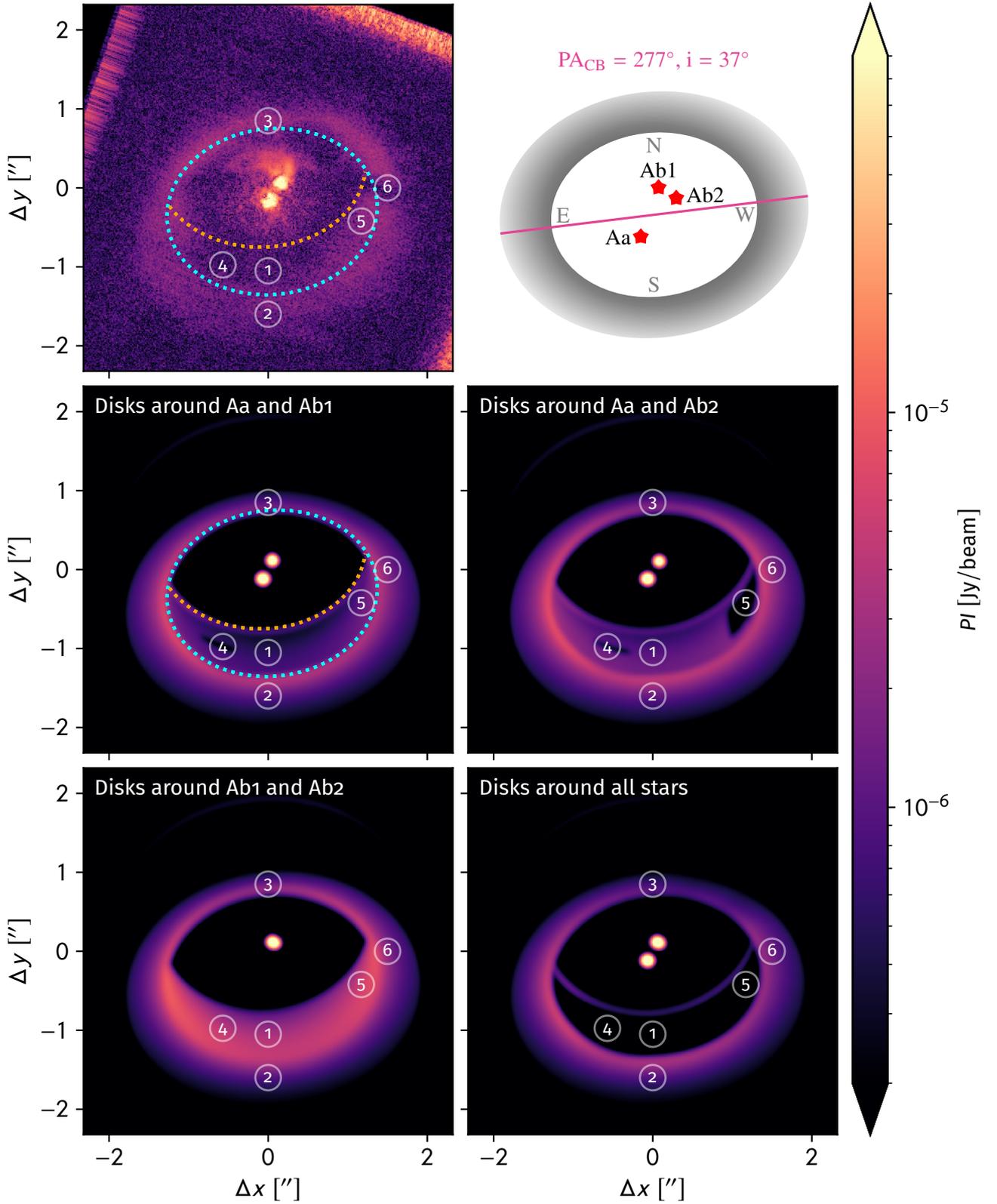}
	\caption{\textit{Top left:} Polarized intensity observation of GG Tau A in the near-infrared (H-Band) with Subaru/HiCIAO \citep{yang_near-infrared_2017}. The uncalibrated observation was fitted visually to the simulated emission maps by assuming a linear photon count. The inner wall of the CB disk is illustrated by ellipses that are fitted to the transition from the inner edge to the upper layers of the CB disk as seen in the simulations (see \textit{center left}). \textit{Other images:} Polarized intensity maps of GG Tau A that are simulated at $\lambda=\SI{1.65}{\micro\metre}$. The images are convolved with a gaussian beam of $\SI{0.07}{\arcsecond}\times\SI{0.07}{\arcsecond}$ to take the resolution of the Subaru/HiCIAO telescope into account. Each simulated map considers a different combination of circumstellar disks around GG Tau Aa, Ab1, and Ab2 to investigate their individual influence on the emission of the circumbinary disk (see Table \ref{tab:cs_disk_configurations}).}
	\label{fig:compare_obs:near_infrared_polarization_2}
\end{figure*}

\begin{figure*}
	\centering
	\includegraphics[width=\hsize]{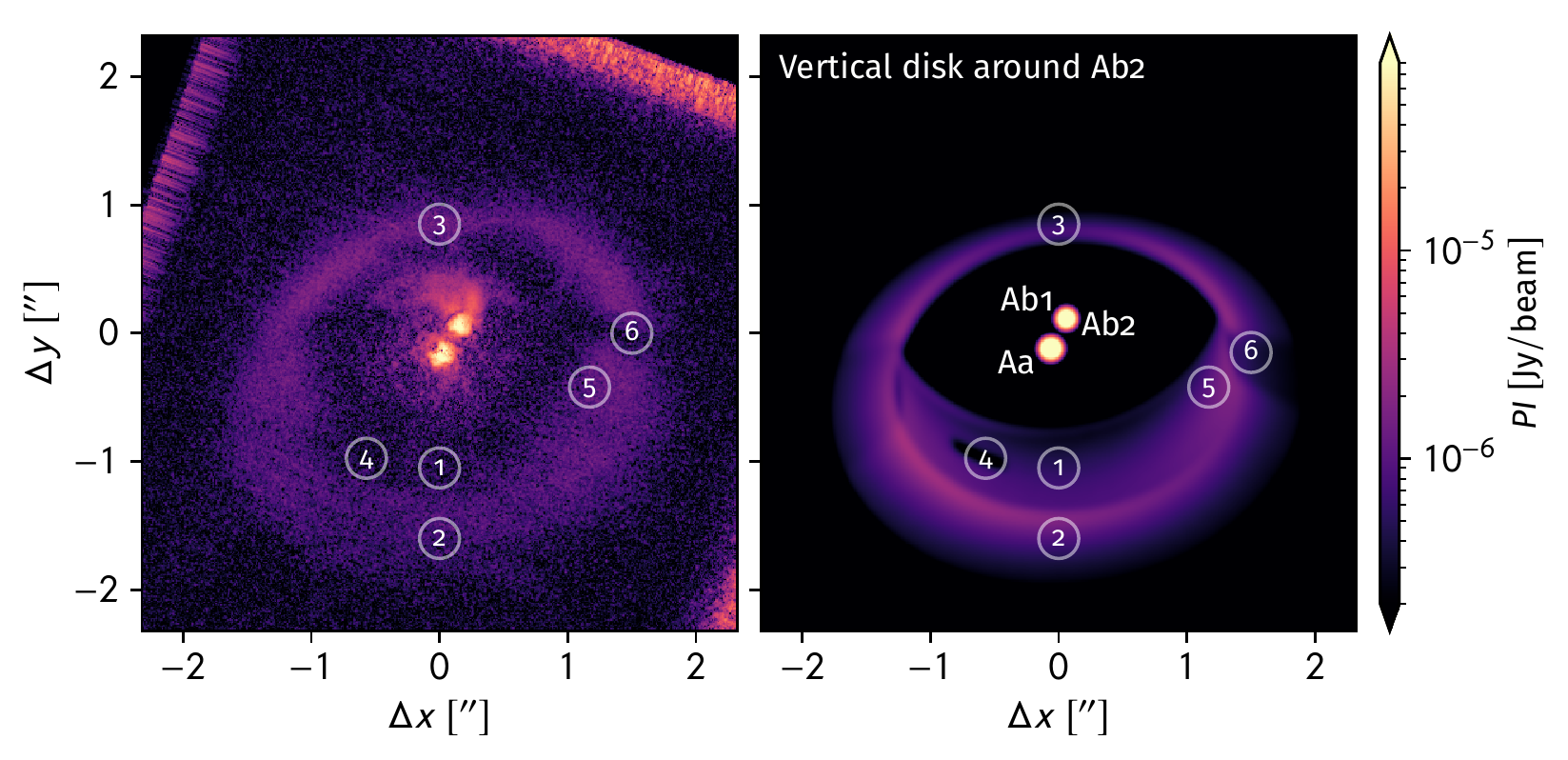}\\
	\caption{\textit{Left:} Polarized intensity observation of GG Tau A in the near-infrared (H-Band) with Subaru/HiCIAO \citep{yang_near-infrared_2017}. The uncalibrated observation was fitted visually to the simulated emission maps by assuming a linear photon count. \textit{Right:} Polarized intensity map of GG Tau A that is simulated at $\lambda=\SI{1.65}{\micro\metre}$. The image is convolved with a gaussian beam of $\SI{0.07}{\arcsecond}\times\SI{0.07}{\arcsecond}$ to take the resolution of the Subaru/HiCIAO telescope into account. The model consists of disks around Aa and Ab1 in the same plane as the circumbinary disk and a disk around Ab2 that is inclined by $\SI{90}{\degree}$ with a position angle of the rotation axis of $\SI{270}{\degree}$ (see Fig. \ref{fig:model_description:circumbinary_disk:orbit_illustration}). An increased grid resolution and sightly larger scale height of the CS disk around Aa was required to achieve the sharp western shadow.}
	\label{fig:compare_obs:near_infrared_polarization_5}
\end{figure*}

In this section, we compare simulated polarized emission maps\footnote{
	In addition to the polarized emission maps, we also simulated intensity maps that we compare to an observation made by Gemini/Hokupa'a (see Figs. \ref{fig:compare_obs:near_infrared_intensity_1} and \ref{fig:compare_obs:near_infrared_intensity_2}). For this observation, we unfortunately have only access to the image and not the data, thus our potential for comparisons is quite low. Nevertheless, we performed the same analysis on the simulated emission maps and found qualitatively the same behavior as we obtained from the polarized emission maps (see Tables \ref{tab:compare_obs:flux_measurements_intensity_1} and \ref{tab:compare_obs:flux_measurements_intensity_2}).}
of our GG Tau A model with a polarimetric observation in the near-infrared made by Subaru/HiCIAO \citep[see Figs. \ref{fig:compare_obs:near_infrared_polarization_1} to \ref{fig:compare_obs:near_infrared_polarization_5},][]{yang_near-infrared_2017}. The simulated emission maps are based on our GG Tau A model in its various configurations as described in Table \ref{tab:cs_disk_configurations} and Sect. \ref{sect:model_description}. With these configurations, we are able to investigate and identify the impact of each CS disk on the emission of the CB disk. To further quantify our results, we define six characteristic regions in the emission maps and obtain their mean emitted (polarized) flux by spatially averaging the emission coming from the corresponding circles in Figs. \ref{fig:compare_obs:near_infrared_polarization_1} to \ref{fig:compare_obs:near_infrared_polarization_5} (circle radius $r_\text{circle}=\SI{0.15}{\arcsecond}$, see Table \ref{tab:compare_obs:flux_measurements_polarization_3}). Regions 2, 3 and 6 are the upper layers of the CB disk, whereas regions 1, 4 and 5 are located on the midplane, in the vertical wall of the CB disk. Regions 4, 5 and 6 represent specific locations of projected shadows by individual CS disks. For each region, we calculate the relative decrease of emission from the configuration without any CS disk to any other configuration (see Table \ref{tab:compare_obs:flux_measurements_polarization_1}). In addition, we present in Table \ref{tab:compare_obs:flux_measurements_polarization_2} for each configuration selected emission ratios between the characteristic regions.

\subsubsection*{Large-scale structures}
The absolute polarized intensity in the midplane at the vertical wall of the CB disk (region 1) can be reduced by a factor of 4 to 10 (and up to 300) depending on the number of CS disks in the system (configurations 2, 5, 8). We are lacking calibrated PI values to make a direct, absolute comparison with the measured intensities, however a first robust conclusion is that two disks around Aa and Ab1 (or Ab2) (configurations 5 and 6) are necessary to reproduce the intensity ratios between the midplane and the upper layers ($\textit{PI}_1 / \textit{PI}_2$ and $\textit{PI}_1 / \textit{PI}_3$). The upper layers of the CB disk are less affected than the midplane: region 2 (most distant layers, rear side of the CB disk) is almost unchanged whatever the configuration, and region 3 (closest to the observer) has intensities reduced by a factor 2 to 4 only. The scale heights of the CS disks have been tuned to maximize the obscuration in the wall of the CB disk without impacting much the most upper layers which still show prominent emission in the observed scattered-light images\footnote{Using different dust properties for the inner disk would result in a different adjustment of the scale heights to obtain the same effect.}. Note that our assumption of a vertical wall for the CB disk (under the hypothesis of hydrostatic equilibrium) has a strong impact on the global shape of the CB disk emission. The observed emission is much smoother, while the vertical wall produces a more contrasted, peaked emission at the upper edge of the CB disk wall. The strong constraint on the CS disks scale height/flaring would be relaxed in the case of a rounded wall for the inner edge of the CB disk. Moreover, hydrodynamic simulations of binary systems show that depending on binary parameters, the height of the CB disk at its inner edge can vary \citep[e.g.,][]{sotnikova_hydrodynamic_2007}. We conclude that the midplane of the CB disk is mostly obscured by the self-shadowing of Aa and Ab1 (or Ab2) disks, and that most of the emission detected in the southern region emanates from the upper layers of the CB disk only.

\subsubsection*{Small-scale structures}
Individual features are also identified in the observations at smaller spatial scales on the CB disk, which our modeling attempts to reproduce:
\begin{enumerate}
    \item Region 4 on the south-eastern side displays a kink in the CB disk wall, which looks like a deeper shadow compared to the symmetric western side. It is approximately aligned with the main binary Aa-Ab direction. In our simulation, it corresponds to the shadow casted by the CS disk around Aa with the source of light being the close-binary Ab1/Ab2. Its exact location and azimuthal extent depend on whether the irradiation comes from Ab1, Ab2, or from both (configurations 2, 5, 6). It appears more contrasted when one of the stars Ab1 or Ab2 has no disk (or if this third disk is inclined with respect to the CB disk midplane, see item 3 below). In our modeling, this shadow is located exactly on the CB disk midplane since the stellar orbits are coplanar with the CB disk, but small ($1$-$\SI{3}{\degree}$) deviations of the orbital inclination could shift it towards upper layers, as seen in the observation.
    \item A similar feature can be seen in region 5 (western side, midplane) in our simulations, which corresponds to the shadow casted by the disk of Ab2, illuminated by the star Ab1. There is no clear counterpart in the observations, however a sharp shadow seems to be projected close to this location on the uppermost layers of the CB disk (region 6).
    \item The tilted dark lane in region 6 looks actually as the prolongation of an east-west dark bar which is aligned with the close-binary Ab1/Ab2, and may extend up to the eastern side of the CB disk (small kink observable at the symmetric location with respect to Ab1/Ab2). Our modeling can roughly reproduce these patterns if one of the CS disks around Ab2 (or Ab1) is significantly tilted with respect to the CB disk midplane (see Fig. \ref{fig:compare_obs:near_infrared_polarization_5}). In this case (configuration 9), the pattern is the self-shadow of Ab2 by its own disk, projected on the upper layers of the disk (its midplane footprint is not visible due to the CB disk inclination and orientation). This so-called vertical disk configuration corresponds to a maximum tilt of $\SI{90}{\degree}$. The location of the sharp western shadow is correctly reproduced in the simulation, but it displays a divergent pattern towards the outer disk. (the symmetric eastern shadow is also present, but is even shallower). In addition, it was necessary to increase the grid resolution and the scale height of the CS disk around Aa to achieve a very sharp western shadow ($h_0=\SI{0.7}{\astronomicalunit}\rightarrow\SI{0.8}{\astronomicalunit}$).
\end{enumerate}
Attributing the sharp shadow pattern on the western side of the CB disk to the self-shadowing by Ab2's own disk is also supported by the absence of rotation of this pattern along the years. Indeed, it appears almost at the same position angle in all optical/near-infrared images since about $\SI{20}{\year}$ \citep{silber_near-infrared_2000, krist_hst/acs_2002, itoh_near-infrared_2002, krist_hubble_2005, itoh_near-infrared_2014, yang_near-infrared_2017}. \cite{itoh_near-infrared_2014} claimed that they detected a counter-clockwise rotation (in the opposite direction to the CB disk rotation) between 2001 and 2011, with an amplitude of about $\SI{5}{\degree}$, which is not further confirmed in the most recent high-contrast images (\citealt{yang_near-infrared_2017}; Keppler et al. in prep). In our scenario, the pattern projected onto the CB disk, would slightly oscillate as Ab2 rotates around the barycenter of the close-binary, with a small amplitude in the order of $\SI{1.4}{\degree}$ (projected separation of about $\SI{4.5}{\astronomicalunit}$ at the distance of $\SI{180}{\astronomicalunit}$). The orbital period of the close-binary star is not yet precisely constrained, but is evaluated in the range $\SI{15}{\year}-\SI{20}{\year}$. If the origin of the dark pattern were a shadow of Ab1 casted by the disk of Ab2 (or reciprocally), this shadow would rotate at the same azimuthal speed as the close-binary itself. In our proposed scenario, only the precession of the disk rotation axis could generate a systematic counter-rotation of the shadow, which usually takes place on timescales longer than the orbital period. Therefore, self-shadowing by a misaligned disk around one of the components of the close-binary star seems to be the most convincing configuration to reproduce the observations. Such misaligned disks in a young binary system have already been reported in the recent years \citep[e.g., HK~Tau, L1551~NE;][]{jensen_misaligned_2014, takakuwa_spiral_2017}. The entire ($\sim\SI{2}{\arcsecond}$ long) east-west dark bar could be attributed to the self-shadowing by the disk of Ab2 projected onto the accretion streams present in the cavity, which feed the inner CS disks with fresh material from the outer CB disk.

\subsection{Implications}
Previous studies usually neglected the impact of the CS disks around the stars of GG Tau A \citep[e.g.,][]{duchene_multiwavelength_2004}. As a result, the southern region of the CB disk is in simulations usually related to the southern inner edge of the CB disk. A sign of this are the wrong distances to the northern and southern region of the CB disk in the simulation compared to the observation \citep[as seen in Fig. 4 and 5 of][]{duchene_multiwavelength_2004}. As a consequence, derived quantities such as the scale height and the dust properties can differ greatly from the real values. For instance, the scale height of the CB disk of $h_0=\SI{21}{\astronomicalunit}$ that we used in our model is in good agreement with the observation of \cite{yang_near-infrared_2017} and much lower than the $h'_0=\SI{32}{\astronomicalunit}$ as used by \cite{duchene_multiwavelength_2004}.

A second important impact of the shadowing by the inner CS disks is the change in the temperature structure of the CB disk. From extreme cases of configurations 1 to 8 in our simulations, the computed temperatures in the disk midplane can be divided by a factor up to 2 at the inner edge of the CB disk (from $T_\text{dust}=\SI{38}{K}$ to $T_\text{dust}=\SI{17}{K}$ at $R=\SI{180}{\astronomicalunit}$). At a larger distance ($R=\SI{250}{\astronomicalunit}$), the reduction is more moderate ($\sim\SI{30}{\percent}$) since the heating is dominated by the stellar irradiation impinging the upper disk layers, which are less affected by the shadowing effect. We note that the calculated midplane dust temperatures of configurations 8 and 9 with strong shadowing are more consistent with the direct determination of the temperature of the mm-sized grains reported by \cite{dutrey_physical_2014} from ALMA and NOEMA data ($T_\text{dust}=\SI{14}{K}$ at $R=\SI{200}{\astronomicalunit}$ and $T_\text{dust}=\SI{11}{K}$ at $R=\SI{250}{\astronomicalunit}$).


\begin{table*}
	\caption{Overview of the emission at different positions in the polarized emission maps (see Figs. \ref{fig:compare_obs:near_infrared_polarization_1} to \ref{fig:compare_obs:near_infrared_polarization_5} and Table \ref{tab:cs_disk_configurations}).}
	\label{tab:compare_obs:flux_measurements_polarization_3}
	\centering
	\begin{tabular}{clcccccc}
		\# & Configuration            & $\textit{PI}_1 [\si{Jy}]$ & $\textit{PI}_2 [\si{Jy}]$ & $\textit{PI}_3 [\si{Jy}]$ & $\textit{PI}_4 [\si{Jy}]$ & $\textit{PI}_5 [\si{Jy}]$ & $\textit{PI}_6 [\si{Jy}]$ \\
		\hline
		1  & No circumstellar disks   & $\SI{4.09e-03}{}$         & $\SI{1.41e-03}{}$         & $\SI{1.64e-03}{}$         & $\SI{5.34e-03}{}$         & $\SI{7.53e-03}{}$         & $\SI{4.06e-03}{}$         \\
		2  & Disk around Aa           & $\SI{1.28e-03}{}$         & $\SI{1.40e-03}{}$         & $\SI{1.15e-03}{}$         & $\SI{1.20e-03}{}$         & $\SI{2.74e-03}{}$         & $\SI{2.93e-03}{}$         \\
		3  & Disk around Ab1          & $\SI{3.28e-03}{}$         & $\SI{1.11e-03}{}$         & $\SI{1.30e-03}{}$         & $\SI{4.32e-03}{}$         & $\SI{5.83e-03}{}$         & $\SI{3.04e-03}{}$         \\
		4  & Disk around Ab2          & $\SI{3.64e-03}{}$         & $\SI{1.25e-03}{}$         & $\SI{1.43e-03}{}$         & $\SI{4.78e-03}{}$         & $\SI{5.08e-03}{}$         & $\SI{3.48e-03}{}$         \\
		5  & Disks around Aa and Ab1  & $\SI{4.63e-04}{}$         & $\SI{1.09e-03}{}$         & $\SI{8.06e-04}{}$         & $\SI{4.36e-04}{}$         & $\SI{1.03e-03}{}$         & $\SI{1.89e-03}{}$         \\
		6  & Disks around Aa and Ab2  & $\SI{8.28e-04}{}$         & $\SI{1.23e-03}{}$         & $\SI{9.41e-04}{}$         & $\SI{7.74e-04}{}$         & $\SI{2.62e-04}{}$         & $\SI{2.34e-03}{}$         \\
		7  & Disks around Ab1 and Ab2 & $\SI{2.84e-03}{}$         & $\SI{9.42e-04}{}$         & $\SI{1.10e-03}{}$         & $\SI{3.76e-03}{}$         & $\SI{4.87e-03}{}$         & $\SI{2.48e-03}{}$         \\
		8  & Disks around all stars   & $\SI{1.29e-05}{}$         & $\SI{9.29e-04}{}$         & $\SI{5.89e-04}{}$         & $\SI{1.54e-05}{}$         & $\SI{5.43e-05}{}$         & $\SI{1.32e-03}{}$         \\
		\hline
		9  & Vertical disk around Ab2 & $\SI{4.65e-04}{}$         & $\SI{8.26e-04}{}$         & $\SI{4.14e-04}{}$         & $\SI{3.70e-04}{}$         & $\SI{9.94e-04}{}$         & $\SI{3.73e-04}{}$         \\
	\end{tabular}
\end{table*}
\begin{table*}
	\caption{Overview of the decrease of emission at different positions in the polarized emission maps (see Figs. \ref{fig:compare_obs:near_infrared_polarization_1} to \ref{fig:compare_obs:near_infrared_polarization_5} and Table \ref{tab:cs_disk_configurations}). The decrease in polarized emission is calculated for each region $i$ and configuration $j$ with $\delta \textit{PI}_i(j)=\frac{\textit{PI}_i(j)-\textit{PI}_i(\text{no disks})}{\textit{PI}_i(\text{no disks})}$.}
	\label{tab:compare_obs:flux_measurements_polarization_1}
	\centering
	\begin{tabular}{clcccccc}
		\# & Configuration            & $\delta \textit{PI}_1 [\si{\percent}]$ & $\delta \textit{PI}_2 [\si{\percent}]$ & $\delta \textit{PI}_3 [\si{\percent}]$ & $\delta \textit{PI}_4 [\si{\percent}]$ & $\delta \textit{PI}_5 [\si{\percent}]$ & $\delta \textit{PI}_6 [\si{\percent}]$ \\
		\hline
		1  & No circumstellar disks   & $\SI{-0}{}$                            & $\SI{-0}{}$                            & $\SI{-0}{}$                            & $\SI{-0}{}$                            & $\SI{-0}{}$                            & $\SI{-0}{}$                            \\
		2  & Disk around Aa           & $\SI{-69}{}$                           & $\SI{-1}{}$                            & $\SI{-30}{}$                           & $\SI{-78}{}$                           & $\SI{-64}{}$                           & $\SI{-28}{}$                           \\
		3  & Disk around Ab1          & $\SI{-20}{}$                           & $\SI{-22}{}$                           & $\SI{-21}{}$                           & $\SI{-19}{}$                           & $\SI{-23}{}$                           & $\SI{-25}{}$                           \\
		4  & Disk around Ab2          & $\SI{-11}{}$                           & $\SI{-12}{}$                           & $\SI{-12}{}$                           & $\SI{-11}{}$                           & $\SI{-33}{}$                           & $\SI{-14}{}$                           \\
		5  & Disks around Aa and Ab1  & $\SI{-89}{}$                           & $\SI{-23}{}$                           & $\SI{-51}{}$                           & $\SI{-92}{}$                           & $\SI{-86}{}$                           & $\SI{-53}{}$                           \\
		6  & Disks around Aa and Ab2  & $\SI{-80}{}$                           & $\SI{-13}{}$                           & $\SI{-43}{}$                           & $\SI{-86}{}$                           & $\SI{-97}{}$                           & $\SI{-42}{}$                           \\
		7  & Disks around Ab1 and Ab2 & $\SI{-31}{}$                           & $\SI{-33}{}$                           & $\SI{-33}{}$                           & $\SI{-30}{}$                           & $\SI{-35}{}$                           & $\SI{-39}{}$                           \\
		8  & Disks around all stars   & $\SI{-100}{}$                          & $\SI{-34}{}$                           & $\SI{-64}{}$                           & $\SI{-100}{}$                          & $\SI{-99}{}$                           & $\SI{-68}{}$                           \\
		\hline
		9  & Vertical disk around Ab2 & $\SI{-89}{}$                           & $\SI{-41}{}$                           & $\SI{-75}{}$                           & $\SI{-93}{}$                           & $\SI{-87}{}$                           & $\SI{-91}{}$                           \\
	\end{tabular}
\end{table*}
\begin{table*}
	\caption{Overview of the ratios between the emission at different positions in the polarized emission maps (see Figs. \ref{fig:compare_obs:near_infrared_polarization_1} to \ref{fig:compare_obs:near_infrared_polarization_5} and Table \ref{tab:cs_disk_configurations}). We obtained the uncertainty of the emission ratios by calculating the standard deviation of the background noise in the observation. In addition, the ratio between the emission of each region and the background noise is shown for the observation at the bottom of the Table (background noise: $\textit{PI}_\text{BG} / \sigma_\text{BG}\sim2$). Each emission ratio that is inside of two sigma of the observed ratio is marked in green (bold) and each other ratio is marked red (normal).}
	\label{tab:compare_obs:flux_measurements_polarization_2}
	\centering
	\begin{tabular}{clcccccc}
		\# & Configuration               & $\textit{PI}_1 / \textit{PI}_2$    & $\textit{PI}_1 / \textit{PI}_3$    & $\textit{PI}_1 / \textit{PI}_4$    & $\textit{PI}_1 / \textit{PI}_5$    & $\textit{PI}_6 / \textit{PI}_3$                                         \\
		\cline{1-7}
		1  & No circumstellar disks      & {\color{red}2.90}                  & {\color{red}2.50}                  & {\color{new_green}\textbf{0.77}}   & {\color{new_green}\textbf{0.54}}   & {\color{red}2.48}                                                       \\
		2  & Disk around Aa              & {\color{new_green}\textbf{0.91}}   & {\color{red}1.12}                  & {\color{new_green}\textbf{1.07}}   & {\color{new_green}\textbf{0.47}}   & {\color{red}2.55}                                                       \\
		3  & Disk around Ab1             & {\color{red}2.96}                  & {\color{red}2.52}                  & {\color{new_green}\textbf{0.76}}   & {\color{new_green}\textbf{0.56}}   & {\color{red}2.33}                                                       \\
		4  & Disk around Ab2             & {\color{red}2.92}                  & {\color{red}2.54}                  & {\color{new_green}\textbf{0.76}}   & {\color{new_green}\textbf{0.72}}   & {\color{red}2.43}                                                       \\
		5  & Disks around Aa and Ab1     & {\color{new_green}\textbf{0.42}}   & {\color{new_green}\textbf{0.57}}   & {\color{new_green}\textbf{1.06}}   & {\color{new_green}\textbf{0.45}}   & {\color{red}2.35}                                                       \\
		6  & Disks around Aa and Ab2     & {\color{new_green}\textbf{0.67}}   & {\color{new_green}\textbf{0.88}}   & {\color{new_green}\textbf{1.07}}   & {\color{red}3.15}                  & {\color{red}2.49}                                                       \\
		7  & Disks around Ab1 and Ab2    & {\color{red}3.01}                  & {\color{red}2.58}                  & {\color{new_green}\textbf{0.75}}   & {\color{new_green}\textbf{0.58}}   & {\color{red}2.26}                                                       \\
		8  & Disks around all stars      & {\color{red}0.01}                  & {\color{red}0.02}                  & {\color{new_green}\textbf{0.84}}   & {\color{new_green}\textbf{0.24}}   & {\color{red}2.24}                                                       \\
		\cline{1-7}
		9  & Vertical disk around Ab2    & {\color{new_green}\textbf{0.56}}   & {\color{red}1.12}                  & {\color{new_green}\textbf{1.26}}   & {\color{new_green}\textbf{0.47}}   & {\color{new_green}\textbf{0.90}}                                        \\
		\cline{1-7}
		10 & Observation (Subaru/HiCIAO) & $0.73\pm0.32$                      & $0.43\pm0.16$                      & $1.20\pm0.65$                      & $0.48\pm0.19$                      & $0.57\pm0.17$                                                           \\
		\\
		\# & Configuration               & $\textit{PI}_1 / \sigma_\text{BG}$ & $\textit{PI}_2 / \sigma_\text{BG}$ & $\textit{PI}_3 / \sigma_\text{BG}$ & $\textit{PI}_4 / \sigma_\text{BG}$ & $\textit{PI}_5 / \sigma_\text{BG}$ & $\textit{PI}_6 / \sigma_\text{BG}$ \\
		\hline
		10 & Observation (Subaru/HiCIAO) & $2.9$                              & $3.9$                              & $6.6$                              & $2.4$                              & $5.9$                              & $3.7$                              \\
	\end{tabular}
\end{table*}

%
\section{Conclusions}
\label{conclusions}
We investigated the multiple star system GG Tau A by using a radiative transfer model that considers not only the CB disk but also the CS disks around the individual stars.

By studying the ratio of polarized intensity at different locations in the CB disk, we conclude that the observed scattered-light near-infrared emission is best reproduced, if the CB disk lies in the shadow of at least two co-planar CS disks surrounding the central stars. This implies that the inner wall of the CB disk is strongly obscured around the midplane, while the observed emission is actually dominated by the most upper disk layers. In addition, the inclined dark lane ("gap") on the western side of the CB disk, which is a stable (non rotating) feature since ${\sim}\SI{20}{\year}$, can only be explained by the self-shadowing of a misaligned CS disk surrounding one of the two components of the secondary close-binary star GG Tau Ab.

From our findings, we conclude that the CS disks around the stars are crucial to understand the emission of multiple star systems in general. This includes the emission at infrared wavelengths coming from the CS disks itself and the decreased scattered-light emission and heating of the CB disk due to the casting of shadows. For instance, the dust temperature derived from mm observations is more consistent with the structure predicted by configurations using at least two co-planar disks. Furthermore, it is important to know the exact configuration to derive constraints on a multiple star system even on large scales. Otherwise, the origin of emission features can be misleading and, therefore, result in wrong conclusions. For instance, the very high scale height that was derived in previous studies for GG Tau A can be explained by interpreting the southern emission as the inner edge of the CB disk instead of its upper layers.

In summary, multiple star systems like GG Tau A are offering complex and diverse environments to study, for instance, the formation of planets. Based on our work, future studies of GG Tau A and multiple star systems in general are able to more accurately analyze observations and radiative transfer simulations by understanding the importance of the CS disks and their influence on the emission of the whole system.

\begin{acknowledgements}
	Part of this work was supported by the \emph{Labex P2IO} project
	number \emph{A-JWST-01-02-01+LABEXP2IOPDO}.

	This research has made use of the VizieR catalogue access tool, CDS,
	Strasbourg, France. The original description of the VizieR service was
	published in A\&AS 143, 23.
\end{acknowledgements}

%
\bibliographystyle{bibtex/aa} 
\bibliography{bibtex/my_library} 
%

\begin{appendix}
	\section{Spectral energy distribution}\label{app:sed}
The black dots in Fig. \ref{fig:comparison_results:sed:all_disks} are photometric measurements obtained from the VizieR catalog service
\citep[used catalogs:][]{de_grijp_warm_1987, mermilliod_ubv_1987, helou_infrared_1988, herbig_third_1988, hauck_general_1990, strauss_redshift_1990, de_grijp_warm_1992, gezari_catalog_1993, urban_ac_1998, saunders_pscz_2000, kharchenko_all-sky_2001, zacharias_naval_2004, zacharias_second_2004, ducourant_pre-main_2005, ammons_n2k_2006, droege_tass_2006, ducourant_pm2000_2006, lawrence_ukirt_2007, di_francesco_scuba_2008,lasker_second-generation_2008, roser_ppm-extended_2008, duchene_planet_2010,ishihara_akari/irc_2010, ita_akaris_2010, roeser_ppmxl_2010, takita_survey_2010, yamamura_vizier_2010, bianchi_galex_2011, fedorov_residual_2011, cutri_vizier_2012, harris_resolved_2012, page_xmm-newton_2012, zacharias_vizier_2012, andrews_mass_2013, mohanty_protoplanetary_2013, cutri_vizier_2014, dias_proper_2014, esplin_wise_2014, toth_akari_2014, abrahamyan_iras_2015, henden_apass_2015, andruk_catalog_2016, chambers_pan-starrs1_2016,  collaboration_gaia_2016, marton_all-sky_2016, nascimbeni_all-sky_2016,altmann_hot_2017, bianchi_revised_2017, huber_vizier_2017, zacharias_vizier_2017}.

\section{Near-infrared observations}\label{app:near_infrared_obs}
\newcommand{\Icaption}{\textit{Top left:} Observation of GG Tau A in the near-infrared (H-Band) with Gemini/Hokupa'a (Credit: Daniel Potter/University of Hawaii Adaptive Optics Group/Gemini Observatory/AURA/NSF). \textit{Other images:} Intensity maps of GG Tau A that are simulated at $\lambda=\SI{1.65}{\micro\metre}$. The images are convolved with a gaussian beam of $\SI{0.05}{\arcsecond}\times\SI{0.05}{\arcsecond}$ to take the resolution of the Gemini/Hokupa'a telescope into account. Each simulated map considers a different combination of circumstellar disks around GG Tau Aa, Ab1, and Ab2 to investigate their individual influence on the emission of the circumbinary disk (see Table \ref{tab:cs_disk_configurations}).}
\begin{figure*}
	\centering
	\input{figures/I_emission_map_illustration_1}
	\caption{\Icaption{}}
	\label{fig:compare_obs:near_infrared_intensity_1}
\end{figure*}
\begin{figure*}
	\centering
	\input{figures/I_emission_map_illustration_2}
	\caption{\Icaption{}}
	\label{fig:compare_obs:near_infrared_intensity_2}
\end{figure*}
\begin{figure*}
	\centering
	\input{figures/I_emission_map_illustration_3}
	\caption{\textit{Left:} Observation of GG Tau A in the near-infrared (H-Band) with Gemini/Hokupa'a (Credit: Daniel Potter/University of Hawaii Adaptive Optics Group/Gemini Observatory/AURA/NSF). \textit{Right:} Intensity maps of GG Tau A that are simulated at $\lambda=\SI{1.65}{\micro\metre}$. The images are convolved with a gaussian beam of $\SI{0.05}{\arcsecond}\times\SI{0.05}{\arcsecond}$ to take the resolution of the Gemini/Hokupa'a telescope into account. The model consists of disks around Aa and Ab1 in the same plane as the circumbinary disk and a disk around Ab2 that is inclined by $\SI{90}{\degree}$ with a position angle of the rotation axis of $\SI{270}{\degree}$ (see Fig. \ref{fig:model_description:circumbinary_disk:orbit_illustration}).}
	\label{fig:compare_obs:near_infrared_intensity_3}
\end{figure*}

\begin{table*}
	\caption{Overview of the emission at different positions in the intensity emission maps (see Figs. \ref{fig:compare_obs:near_infrared_intensity_1} and \ref{fig:compare_obs:near_infrared_intensity_2} as well as Table \ref{tab:cs_disk_configurations}).}
	\label{tab:compare_obs:flux_measurements_intensity_3}
	\centering
	\begin{tabular}{clcccccc}
		\# & Configuration            & $F_1 [\si{Jy}]$   & $F_2 [\si{Jy}]$   & $F_3 [\si{Jy}]$   & $F_4 [\si{Jy}]$   & $F_5 [\si{Jy}]$   & $F_6 [\si{Jy}]$   \\
		\hline
		1  & No circumstellar disks   & $\SI{8.93e-03}{}$ & $\SI{1.92e-03}{}$ & $\SI{9.76e-03}{}$ & $\SI{9.57e-03}{}$ & $\SI{1.04e-02}{}$ & $\SI{4.86e-03}{}$ \\
		2  & Disk around Aa           & $\SI{2.76e-03}{}$ & $\SI{1.91e-03}{}$ & $\SI{6.91e-03}{}$ & $\SI{2.15e-03}{}$ & $\SI{4.02e-03}{}$ & $\SI{3.69e-03}{}$ \\
		3  & Disk around Ab1          & $\SI{7.16e-03}{}$ & $\SI{1.50e-03}{}$ & $\SI{7.42e-03}{}$ & $\SI{7.73e-03}{}$ & $\SI{7.89e-03}{}$ & $\SI{3.72e-03}{}$ \\
		4  & Disk around Ab2          & $\SI{7.96e-03}{}$ & $\SI{1.69e-03}{}$ & $\SI{8.37e-03}{}$ & $\SI{8.58e-03}{}$ & $\SI{6.75e-03}{}$ & $\SI{4.25e-03}{}$ \\
		5  & Disks around Aa and Ab1  & $\SI{9.96e-04}{}$ & $\SI{1.48e-03}{}$ & $\SI{4.57e-03}{}$ & $\SI{7.75e-04}{}$ & $\SI{1.50e-03}{}$ & $\SI{2.54e-03}{}$ \\
		6  & Disks around Aa and Ab2  & $\SI{1.79e-03}{}$ & $\SI{1.68e-03}{}$ & $\SI{5.52e-03}{}$ & $\SI{1.40e-03}{}$ & $\SI{3.52e-04}{}$ & $\SI{3.07e-03}{}$ \\
		7  & Disks around Ab1 and Ab2 & $\SI{6.20e-03}{}$ & $\SI{1.27e-03}{}$ & $\SI{6.03e-03}{}$ & $\SI{6.74e-03}{}$ & $\SI{6.47e-03}{}$ & $\SI{3.10e-03}{}$ \\
		8  & Disks around all stars   & $\SI{3.56e-05}{}$ & $\SI{1.25e-03}{}$ & $\SI{3.17e-03}{}$ & $\SI{3.78e-05}{}$ & $\SI{6.99e-05}{}$ & $\SI{1.92e-03}{}$ \\
		\hline
		9  & Vertical disk around Ab2 & $\SI{9.93e-04}{}$ & $\SI{1.14e-03}{}$ & $\SI{2.69e-03}{}$ & $\SI{6.52e-04}{}$ & $\SI{1.47e-03}{}$ & $\SI{8.23e-04}{}$ \\
	\end{tabular}
\end{table*}
\begin{table*}
	\caption{Overview of the decrease of emission at different positions in the intensity emission maps (see Figs. \ref{fig:compare_obs:near_infrared_intensity_1} and \ref{fig:compare_obs:near_infrared_intensity_2} as well as Table \ref{tab:cs_disk_configurations}). The decrease in emission is calculated for each region $i$ and configuration $j$ with $\delta F_i(j)=\frac{F_i(j)-F_i(\text{no disks})}{F_i(\text{no disks})}$.}
	\label{tab:compare_obs:flux_measurements_intensity_1}
	\centering
	\begin{tabular}{clcccccc}
		\# & Configuration            & $\delta F_1 [\si{\percent}]$ & $\delta F_2 [\si{\percent}]$ & $\delta F_3 [\si{\percent}]$ & $\delta F_4 [\si{\percent}]$ & $\delta F_5 [\si{\percent}]$ & $\delta F_6 [\si{\percent}]$ \\
		\hline
		1  & No circumstellar disks   & $\SI{-0}{}$                  & $\SI{-0}{}$                  & $\SI{-0}{}$                  & $\SI{-0}{}$                  & $\SI{-0}{}$                  & $\SI{-0}{}$                  \\
		2  & Disk around Aa           & $\SI{-69}{}$                 & $\SI{-1}{}$                  & $\SI{-29}{}$                 & $\SI{-78}{}$                 & $\SI{-61}{}$                 & $\SI{-24}{}$                 \\
		3  & Disk around Ab1          & $\SI{-20}{}$                 & $\SI{-22}{}$                 & $\SI{-24}{}$                 & $\SI{-19}{}$                 & $\SI{-24}{}$                 & $\SI{-24}{}$                 \\
		4  & Disk around Ab2          & $\SI{-11}{}$                 & $\SI{-12}{}$                 & $\SI{-14}{}$                 & $\SI{-10}{}$                 & $\SI{-35}{}$                 & $\SI{-13}{}$                 \\
		5  & Disks around Aa and Ab1  & $\SI{-89}{}$                 & $\SI{-23}{}$                 & $\SI{-53}{}$                 & $\SI{-92}{}$                 & $\SI{-86}{}$                 & $\SI{-48}{}$                 \\
		6  & Disks around Aa and Ab2  & $\SI{-80}{}$                 & $\SI{-13}{}$                 & $\SI{-43}{}$                 & $\SI{-85}{}$                 & $\SI{-97}{}$                 & $\SI{-37}{}$                 \\
		7  & Disks around Ab1 and Ab2 & $\SI{-31}{}$                 & $\SI{-34}{}$                 & $\SI{-38}{}$                 & $\SI{-30}{}$                 & $\SI{-38}{}$                 & $\SI{-36}{}$                 \\
		8  & Disks around all stars   & $\SI{-100}{}$                & $\SI{-35}{}$                 & $\SI{-68}{}$                 & $\SI{-100}{}$                & $\SI{-99}{}$                 & $\SI{-61}{}$                 \\
		\hline
		9  & Vertical disk around Ab2 & $\SI{-89}{}$                 & $\SI{-41}{}$                 & $\SI{-73}{}$                 & $\SI{-93}{}$                 & $\SI{-86}{}$                 & $\SI{-83}{}$                 \\
	\end{tabular}

\end{table*}
\begin{table*}
	\caption{Overview of the ratios between the emission at different positions in the intensity emission maps (see Figs. \ref{fig:compare_obs:near_infrared_intensity_1} and \ref{fig:compare_obs:near_infrared_intensity_2} as well as Table \ref{tab:cs_disk_configurations}).}
	\label{tab:compare_obs:flux_measurements_intensity_2}
	\centering
	\begin{tabular}{clccccc}
		\# & Configuration            & $F_1 / F_2$ & $F_1 / F_3$ & $F_1 / F_4$ & $F_1 / F_5$ & $F_6 / F_3$ \\
		\hline
		1  & No circumstellar disks   & 4.64        & 0.92        & 0.93        & 0.86        & 0.50        \\
		2  & Disk around Aa           & 1.45        & 0.40        & 1.29        & 0.69        & 0.53        \\
		3  & Disk around Ab1          & 4.79        & 0.97        & 0.93        & 0.91        & 0.50        \\
		4  & Disk around Ab2          & 4.70        & 0.95        & 0.93        & 1.18        & 0.51        \\
		5  & Disks around Aa and Ab1  & 0.67        & 0.22        & 1.28        & 0.66        & 0.56        \\
		6  & Disks around Aa and Ab2  & 1.07        & 0.32        & 1.28        & 5.09        & 0.56        \\
		7  & Disks around Ab1 and Ab2 & 4.89        & 1.03        & 0.92        & 0.96        & 0.51        \\
		8  & Disks around all stars   & 0.03        & 0.01        & 0.94        & 0.51        & 0.61        \\
		\hline
		9  & Vertical disk around Ab2 & 0.87        & 0.37        & 1.52        & 0.67        & 0.31        \\
	\end{tabular}
\end{table*}
\end{appendix}

\end{document}